
\input harvmac
\def\np#1#2#3{Nucl. Phys. B{#1} (#2) #3}
\def\pl#1#2#3{Phys. Lett. {#1}B (#2) #3}

\def\physrev#1#2#3{Phys. Rev. {D#1} (#2) #3}

\def\prep#1#2#3{Phys. Rep. {#1} (#2) #3}

\def\pr#1#2#3{Phys. Rev. D{#1} (#2) #3}
\def\ev#1{\langle#1\rangle}
\def\pf{{\rm Pf ~}}
\def\tilde{\widetilde}

\Title{hep-th/9507018, RU-95-46}
{\vbox{\centerline{Chiral Duals of Non-Chiral SUSY Gauge Theories}}}
\bigskip
\centerline{P. Pouliot}
\vglue .5cm
\centerline{Department of Physics and Astronomy}
\centerline{Rutgers University}
\centerline{Piscataway, NJ 08855-0849, USA}

\bigskip

\noindent

We study $N=1$ SUSY gauge theories in four dimensions with gauge group
$Spin(7)$ and $N_f$ flavors of
quarks in the spinorial representation.
We find that in the range $6< N_f < 15$, this theory has a
long distance description in terms of an $SU(N_f-4)$
gauge theory with a symmetric tensor and $N_f$ antifundamentals.
As a spin-off, we obtain by deforming along a flat direction
a dual description of
the theories based on the exceptional gauge group $G_2$ with $N_f$
fundamental flavors of quarks.

\Date{6/95}

\nref\powerh{N. Seiberg, the Power of Holomorphy -- Exact
Results in 4D SUSY Field Theories, Proc. of PASCOS 94, hep-th/9408013,
 RU-94-64, IASSNS-HEP-94/57}
\nref\powerd{N. Seiberg, the Power of Duality
 -- Exact Results in 4D SUSY Field Theories,
Proc. of PASCOS 95 and Proc. of the
Oskar Klein Lectures, hep-th/9506077, RU-95-37, IASSNS-HEP-95/46}
\nref\otherads{I. Affleck, M. Dine and N. Seiberg, \np{241}{1984}{493}}
\nref\ads{I. Affleck, M. Dine and N. Seiberg,
\np{256}{1985}{557}}%
\nref\rus{A.I. Vainshtein, V.I. Zakharov and M.A. Shifman,
Sov. Phys. Usp. 28 (1985) 709}
\nref\cern{D. Amati,
K. Konishi, Y. Meurice, G.C. Rossi and G.
Veneziano, \prep{162}{1988}{169} }
\nref\sem{N. Seiberg, \np{435}{1995}{129}}%
\nref\olive{C. Montonen and D. Olive, \pl{72}{1977}{117}}
\nref\switten{N. Seiberg and E. Witten, \np{426}{1994}{19};
 \np{431}{1994}{484}}

Over the past two years, it has become clear that supersymmetry renders
possible the precise study of dynamical phenomena in
quantum field theory.
For recent reviews and lists of references, see \refs{\powerh,\powerd};
for reviews to earlier work, see \refs{\otherads - \cern}.
The outstanding idea in understanding $N=1$ SUSY
theories is Duality~\sem: a theory which is in a non-Abelian
Coulomb phase at large distances may
have  an equivalent description in terms of a different
gauge theory, with different gauge group and matter content.
A complicated, strongly coupled theory might have a weakly coupled dual,
and that allows the theory to be solved.
This duality is a generalization of the Montonen-Olive duality
\refs{\olive,\switten} of extended supersymmetry.

\nref\son{K. Intriligator and N. Seiberg, hep-th/9503179,
RU-95-3, IASSNS-HEP-95/5; hep-th/9506084,
RU-95-40, IASSNS-HEP-95/48}
\nref\intpou{K. Intriligator and P. Pouliot, hep-th/9505006, RU-95-23}
\nref\kut{D. Kutasov, hep-th/9503086, EFI-95-11}
\nref\ofer{O. Aharony, J. Sonnenschein and S. Yankielowicz,
hep-th/9504113, TAUP-2246-95, CERN-TH/95-91}
\nref\kuschwim{ D. Kutasov and A. Schwimmer,
hep-th/9505004, EFI-95-20, WIS/4/95}
\nref\berk{M. Berkooz, hep-th/9505067, RU-95-29}
\nref\intril{K. Intriligator, hep-th/9505051, RU-95-27}
\nref\leighstr{R.G. Leigh
and M.J. Strassler, hep-th/9505088, RU-95-30}
\nref\ilstr{K. Intriligator, R.G. Leigh and M.J. Strassler,
hep-th/9506148, RU-95-38}
\nref\nonren{N. Seiberg, \pl{318}{1993}{469}}%
\nref\nativacua{N. Seiberg, \pr{49}{1994}{6857}}

In attempting to better understand the nature and generality of this
$N=1$ non-Abelian duality,
a number of examples have been found. One class of examples consists of
 $SU(N)$, $SO(N)$ and $Sp(N)$ gauge theories
with matter in the fundamental representations
originally found in \sem\ and further
studied in \refs{\sem,\son, \intpou}.
In a second class of examples, one considers a theory
under some perturbation
of the superpotential \kut. Its effect
is to constrain the R-symmetry and
to reduce the number of independent chiral operators.
The number of examples of this type
found so far is impressive \refs{\kut - \ilstr},
giving support to the idea that duality is the norm rather
than a curiosity, at least for supersymmetric theories~\powerd.

In this letter, we study new examples of duality,
without requiring a perturbation by
a superpotential.
We encounter the situation
that vector-like models have chiral duals.
In a chiral supersymmetric gauge theory, the chiral superfields
(seen as a large reducible representation) do not fall in a real
representation of the gauge group.
Examples of duality in chiral models have been found in
\refs{\berk, \ilstr}. There, the chiral theories had chiral duals.
Nevertheless, there is nothing wrong a priori  in a non-chiral theory
having a chiral dual.
 As in \sem, we are unable to prove rigorously that
the theories we present are dual. However, we give a number of
arguments that strongly support it to be the case.
The presentation of our results
follows closely that of Seiberg's seminal paper \sem.

The theory that we study is based on the gauge group $Spin(7)$, with
$N_f$ fundamental flavors of quarks, $Q_f$, $f=1,\ldots, N_f$.
This group is 21 dimensional and the adjoint
representation has Casimir index 10;
its spinorial representation
is real and 8 dimensional and its Casimir index is 2.
We refer to this theory as the `electric' theory.
The continuous global symmetry is $SU(N_f)\times U(1)_R$,
where $U(1)_R$ is an R-symmetry.
The quantum numbers of the superfields are listed below:
\eqn\spinmodel{\matrix{
& Spin(7) & SU(N_f) & U(1)_R \cr \cr
Q & 8 & N_f & 1 -{5\over N_f}. \cr} }

\nref\ils{K. Intriligator, R.G. Leigh and N. Seiberg,
\physrev{50}{1994}{1092}}
\nref\ed{E. Witten, unpublished}
\nref\pesando{I. Pesando, hep-th/9506139, NORDITA-95/42 P}
\nref\giddings{S.B. Giddings and J.M. Pierre, hep-th/9506196,
UCSBTH-95-14}

The behavior of this theory for $N_f\le 6$ is standard
\refs{\otherads, \nonren,
 \nativacua}, so we will only briefly state the results.
The independent gauge
invariant operators built out of $Q_f$
 are a symmetric tensor $M=Q^2$ (mesons), and a
totally antisymmetric 4-index
tensor $B=Q^4$ (baryons, for $N_f\ge 4$).
Classically, there are exactly flat directions of the potential,
labeled by the expectation value of $Q_f$, and
along which the theory follows a
\eqn\chain{Spin(7)\to G_2\to SU(3)\to SU(2) \to 1}
pattern of symmetry breaking.
Using holomorphy and the symmetries, we find that
for $N_f\le 3$, a superpotential is generated dynamically:
\eqn\dynsuper{W_{dyn}= \left({\Lambda^{15-N_f} \over \det M}
\right)^{1/( 5-N_f)},}
where $\Lambda$ is the dynamical scale of the theory.
For $N_f=4$ however, the superpotential is determined
by holomorphy and the symmetries only up to some function of
$\det M/B^2$. Observing that when
$\det M- B^2=0$, an $SU(2)$ subgroup
of $Spin(7)$ remains unbroken classically,
it follows by the techniques of
 \ils\ that the superpotential has a pole and thus takes the
form: \eqn\dynsupermore{
{\Lambda^{11} \over \det M -B^2} }
(in this expression
 and elsewhere, for the sake of clarity,
we do not write explicitly the numerical coefficients
(e.g. permutation symmetry factors) but it is
a straightforward matter to determine them).
This superpotential is generated by instantons while
 \dynsuper\ is generated by gaugino condensation (for $N_f\le 3$). These
theories do not have a ground state. For $N_f\ge 5$, there
is no dynamically generated superpotential.
For $N_f=5$, the fields are classically
constrained by the equation
$\det M - M_{ij} B^i B^j=0$, where
$B^i=\epsilon^{ijklm} B_{jklm}$;
however, this constraint is modified by instanton effects to
$\det M - M_{ij} B^i B^j=\Lambda^{10}$.
For $N_f\ge 6$, the quantum moduli space
 of vacua is the same as the classical one.
This can be seen by turning on a mass term
$\tr\ mM$ and adding it to the expression \dynsuper.
The equations
of motion imply
$\ev{M_{ij}} = ({\Lambda^{15-N_f}
\det m})^{1/ 5} (m^{-1})_{ij}$ (and
$\ev B=0$),
so that by taking various limits $m\to 0$, all classical values of
$M$ with $B=0$ can be obtained.
For $N_f=6$, the expectation values of $M$ and $B$
are constrained by
\eqn\nfsixconstraints{\det M (M^{-1})^{ij} -
M_{kl} B^{ik} B^{jl} = 0
\qquad {\rm and}\qquad M_{ik} M_{jl} B^{kl}
+\epsilon_{ijklmn} B^{kl} B^{mn} = 0,}
where $B^{ij}= \epsilon^{ijklmn} B_{klmn}$.
It can be checked that these constraints
properly reduce along the $SU(3)$ flat direction
 to the constraints on the mesons and
baryons
of $SU(3)$ with 4 fundamental flavors \nativacua.
Quantum mechanically however, the theory
confines and is described by
the unconstrained fields $M$ and $B$
with a superpotential proportional to
\eqn\sixsuper{
\det M - M_{ik} M_{jl} B^{ij} B^{kl}
- \pf{B},}
whose equations of motion reproduce
the constraints \nfsixconstraints.
This effective theory of mesons and baryons
satisfies the 't Hooft anomaly matching
conditions at the origin $M=B=0$
where the global symmetry is unbroken.

For $7\le N_f \le 14$,
the theory at the origin of the moduli space is
in an interacting non-Abelian Coulomb phase:
the degrees of freedom are the electric quarks and gluons in
the infrared. This is clear since for $N_f\ge 7$,
there is a flat direction along
which the group is broken to $SU(3)$ with
$5\le N_f-2$ flavors remaining and these theories are known
\refs{\nativacua,\sem} to be
in a non-Abelian Coulomb phase.
The main claim of this paper is that in the range
$7\le N_f \le 14$, the electric theory described above
has an equivalent description in the extreme infrared
 in terms of a different, `dual'
gauge theory. We refer to this
theory as magnetic and without further
ado, state its content.
It is an $SU(N_f-4)$ gauge theory
with a symmetric tensor $s$, antifundamental
quarks $q^f$, $f=1,\ldots,N_f$
 and mesons $M$,
whose transformation properties
under the gauge and global symmetries are:
\eqn\spindual{\matrix{
& SU(N_f-4) & SU(N_f) & U(1)_R  \cr \cr
q & \overline{N_f-4} & \overline{N_f}
& {5\over N_f} - {1\over N_f-4} \cr
s & {(N_f-4)(N_f-3)\over 2} & 1 & {2\over N_f-4} \cr
M & 1 & {N_f (N_f+1)\over 2} & 2-{10\over N_f}. \cr} }
These fields have interactions dictated by the superpotential
\eqn\spinsup{ {1\over \mu^2} M_{fg} q^{fa} q^{gb} s_{ab} +
{1\over \mu^{N_f-7}} \det s }
where $\mu$ is a dimensionful parameter that is required to give the
superpotential a dimension of 3, since we take the dimension of
$M$ to be the one at the ultraviolet fixed point
of the electric theory \son.
{}From this point on, we will set $\mu=1$.
Using the techniques of \nonren,
it follows that the unperturbed
magnetic superpotential does not receive quantum corrections.
As in \sem, we are unable to prove
that this magnetic theory is dual to
the $Spin(7)$ theory with $N_f$
flavors; however, we provide below a
number of consistency checks,
the sum of which
we consider to be strong evidence for duality.

The $U(1)_R$ symmetry so defined is anomaly free.
 Moreover the 't Hooft anomaly matching
conditions are satisfied. They are,
in both the electric and magnetic theories:
$SU(N_f)^3:\  8\ ;
SU(N_f)^2 U(1)_R: \ {-40\over N_f}\ ;
U(1)_R: \ -19\ ;
U(1)_R^3: \ 21- {1000\over N_f^2}$.

Although one cannot talk about the
particle spectrum of such interacting conformal
field theories, there should be a one-to-one correspondence that
preserves the symmetries
between the gauge invariant operators of the electric and magnetic
 theories.
$Q^2$ is naturally mapped to $M$ and
$B=Q^4$ is mapped to $b=q^{N_f-4}$. Under this mapping,
the symmetries are preserved.
The gauge invariant operators $q^2s$ and $\det s$
are redundant and vanish
identically by the equations of motion.

For $N_f\ge 15$, the theory is not asymptotically
free, therefore it is a free theory of gluons and quarks
in the infrared. The magnetic description should not
be valid there. To see this, note that when $N_f\ge 15$, the gauge
invariant operator $q q s$ has R-charge less than $2/3$.
By unitarity, it must be a free field if
the magnetic theory is in a non-Abelian Coulomb phase.
But there is certainly no field with such symmetry
properties in the free electric theory,
which confirms that the magnetic theory is very strongly
coupled for $N_f\ge 15$.
For $N_f=7$, the R-charge of the
meson $M$ is $4/7<2/3$, and thus must
be a free field so that the theory be unitary; however the whole
theory is not free.
There is no value of $N_f$ for
which the theory is in a free magnetic phase.
One might hope that this fact,
which
 makes the analysis more difficult,
is not a generic feature of chiral theories.
Later we will give an example with a free magnetic description.

We now study the mass perturbations of the $Spin(7)$ theory.
Consider giving a mass $m M_{N_f,N_f}$
to the last flavor of the $Spin(7)$ theory
with $N_f$ flavors, $14 \ge N_f\ge 8$.
In the infrared, the theory flows to a
 $Spin(7)$ theory with
$N_f-1$ flavors. The effect of this perturbation
on the magnetic $SU(N_f-4)$ theory with $N_f$ flavors
can be found by studying the
equations of motion following {}from the superpotential
$m M_{N_f,N_f}+ M q^2 s + s^{N_f-4}$.
The equation of motion for $M_{N_f,N_f}$ is
$\ev{q^{N_f} q^{N_f} s} = -m$.
Thus $q^{N_f}$ and $s$ acquire expectation values. By a
gauge transformation, we can
take $\ev{q^{N_f}}=(0,\ldots,0,q)$, with
 $\ev{s_{N_f-4,N_f-4}} \neq 0$ and all other VEVs of
$M$, $s$ and $q$ vanishing.
It is clear that all the equations of motion are then satisfied.
Thus the magnetic group is higgsed to $SU(N_f-5)$.
Along the way, $2N_f-9$ fields have been eaten
by the super-Higgs mechanism.
They are the $2N_f-8$ fields $s_{N_f-4,i}$ and $q^{N_f,i}$,
$i=1, \ldots N_f-4$, {}from which one field is subtracted because
of the constraint $\ev{q^{N_f} q^{N_f} s} = -m$.
The superpotential becomes
$M q^2 s +  s^{N_f-5}$, where now the (suppressed)
flavor indices
run {}from $1$ to $N_f-1$ and a scale
$\ev{s_{N_f-4,N_f-4}}$ has been absorbed by a redefinition of $s$ and $q$.
This perturbed magnetic
theory is just an $SU(N_f-5)$ with $N_f-1$ flavors and with the
appropriate superpotential to be precisely
the dual of the $Spin(7)$ theory with $N_f-1$ flavors.

A more careful analysis is required when $N_f=7$. Giving a mass
$m M_{77}$ to the electric quarks, the magnetic theory
is higgsed to $SU(2)$, with 6 doublets (3 flavors) $\hat q^{if}$
  and a triplet $\hat s^{ij}$
and the mesons $\hat M_{fg}$ ($f,g=1,\ldots 6$; $i,j =1,2$).
The dual superpotential becomes
$\hat s^2 + \hat M \hat q \hat q \hat s$.
The field $\hat s$ is massive and should
be integrated out. The resulting $SU(2)$ theory with 3 flavors
is known to confine \nativacua, and to consist of a theory
of mesons $\hat b^{fg} = \hat q^{fi}
\hat q^{gj} \epsilon_{ij}$. Therefore the dual
superpotential should be rewritten in terms of these meson fields
 and of $\hat M_{fg}$. The strong $SU(2)$ dynamics
produces a superpotential $\pf{\hat b}$.
The result of integrating out $\hat s$ is a contribution
$(\hat M_{fg} \hat q^{fi} \hat q^{gj})
(\hat M_{f'g'} \hat q^{f'i} \hat q^{g'j})
= \hat M_{fg} \hat M_{f'g'} \hat b^{ff'} \hat b^{gg'}$.
We also add a new contribution
$\det \hat M$, which is consistent with the symmetries.
We have not identified the dynamical mechanism generating it.
After identifying the mesons $\hat b$ to the baryons $B$ of the
confining $Spin(7)$, $N_f=6$  theory,
the superpotential that results is
$\det M - M M B B - \pf{ B}$
which is the correct superpotential \sixsuper.
Therefore, by integrating out
a quark {}from the $Spin(7)$, $N_f=7$ theory,
both the electric and magnetic
theories flow to the confining $Spin(7)$,
$N_f=6$ theory. And then, by giving masses and integrating out quarks
{}from this $N_f=6$ theory,
all the results
listed above for $N_f< 6$ are recovered.

We now study a dual pair which follows
{}from our results for $Spin(7)$.
We consider taking as
the electric theory a theory with
gauge group $SU(N_f-4)=SU(N_c)$, $N_c\ge 3$,
 and with $N_f$ fundamentals $Q$ and a
conjugate symmetric tensor $S$.
Along the lines of \kut, we add
a superpotential $\det S$ to this
electric theory. Its field content and symmetries are
\eqn\chiralelectric{\matrix{
& SU(N_c) & SU(N_c+4) & U(1)_R \cr \cr
Q & N_c & N_c+4 & {5\over N_c+4} - {1\over N_c} \cr
S & \overline{N_c(N_c+1)\over 2}
& 1 & {2\over N_c}. \cr} }
It turns out that this theory is much easier to analyze than when
$Spin(7)$ was the electric theory.
The magnetic theory is trivially determined to be:
\eqn\spinsevenmag{\matrix{
& Spin(7) & SU(N_c+4) & U(1)_R \cr \cr
q & 8 & \overline{N_c+4} & 1-{5\over N_c+4} \cr
M &  1 & {(N_c+4)(N_c+5)\over 2} & {10\over N_c+4}, \cr } }
with the simple superpotential $M q q$.
The R-symmetry is anomaly free and the 't
Hooft anomaly matching conditions
are satisfied. The gauge invariant
chiral operators are in one-to-one correspondence:
$M=QQS\to M$, $Q^{N_c}\to q^4$; $qq$ and $\det S$ are redundant.
In both theories, there is no quantum corrections
to the unperturbed superpotentials.
When $N_c\ge 11$, we find that this electric theory with a symmetric
tensor and fundamentals
is at very strong coupling, but that
it has a magnetic description which is free.

To analyze the flat directions of the electric theory, it is
convenient to use the most general solution
of the $D$-terms for $S$ and $Q$
found in \ads:
\eqn\adssolution{
Q_{if} = \pmatrix{Q_1 \cr
& Q_2 \cr
& & \cdots \cr
& & & Q_{N_c} & & \cr} \qquad
S = {\rm diag}(S_1, \ldots, S_{N_c}) }
where $|Q_i|^2-|S_i|^2 = {\rm constant}$ for $i=1,\ldots, N_c$.
This is easily seen by first diagonalizing
the hermitian matrix $S^{\dag ik}S_{kj}$.

It is necessary for consistency that the flat directions of the
dual theories be precisely the same.
{}From the equation of motion for $S$, we
find that $S$, and thus $M=QQS$, have rank at most $N_c-2$
in the electric theory.
It is easy to see that $M$ in the magnetic theory also does not have
rank larger than $N_c-2$. If its rank were larger than $N_c-1$,
the dual $Spin(7)$ theory, after integrating out the massive
dual quarks, would have less than $5$
dual quarks remaining. This $Spin(7)$
theory generates the superpotential \dynsuper\ which removes all the
vacua. If its rank were $N_c-1$, the $Spin(7)$ theory with 5 remaining
dual quarks would have to obey the constraint $\det N - N b b=
\tilde\Lambda^{10}$, with $N=qq$ and $b=q^4$; however, $N=0$ by the
equation of motion for $M$. If the rank of $M$ is less than $N_c-1$,
the dual $Spin(7)$ theory keeps its classical flat directions as
explained earlier. This shows that the mesonic flat directions are
the same in the electric and magnetic theories.
To summarize this example, a chiral theory with a superpotential
has a simpler, vector-like dual.

We consider now the deformation of the $Spin(7)$ electric theory along
its $G_2$ flat direction. Essential results on the group theory of $G_2$
will be obtained with little work. We only need to know that $G_2$ is
the subgroup of $Spin(7)$ left unbroken when a spinor gets a VEV, that
its fundamental representation is $7$ dimensional
and its adjoint $14$ dimensional.
 Starting with $Spin(7)$ with $N_f+1$ flavors,
$Q_0$ and $Q_i$, $i=1,\ldots,N_f$,
consider giving an expectation value to the meson $\ev{M_{00}} \neq 0$,
while other $M$ have vanishing expectation value. The electric theory
is $G_2$ with $N_f$ flavors of quarks $Q$, in the 7 dimensional
fundamental representation of $G_2$ and
transforming as
\eqn\gtwomodel{\matrix{
& G_2 & SU(N_f) & U(1)_R  \cr \cr
Q & 7 & N_f & 1-{4\over N_f} . \cr} }
Since the spinor decomposes as $8=1+7$ and the adjoint
as $21=14+7$, the Casimir
index of the $7$ of $G_2$ is 2, while that of the
$14$ of $G_2$ is $10-2=8$. Thus
this model is asymptotically free for $N_f<12$.
We now study the effect of the $\ev {M_{00}}$ perturbation on the dual.
{}From the equation of motion for $q^i$, $M_{i0}=0$.
The superpotential becomes
\eqn\dualsupgtwo{q^0 q^0 s
+ s^{N_f-3} + M qq s,}
where we have absorbed the scale $\ev{M_{00}}$
in the definition of $q^0\equiv q^{N_f+1}$.
Thus, the dual of $G_2$ with $N_f$ flavors consists of the fields
$q$, $q^0$, $s$ and $M$ transforming as
\eqn\gtwodual{\matrix{
& SU(N_f-3) & SU(N_f) & U(1)_R \cr \cr
q & \overline{N_f-3} &
\overline{ N_f} & {3\over N_f} (1-{1\over N_f-3}) \cr
q^0 & \overline{N_f-3} & 1 & {1- {1\over N_f-3} } \cr
s & {(N_f-3)(N_f-2)\over 2} & 1 & {2\over N_f-3} \cr
M & 1 & {N_f(N_f+1)\over 2} & {2- {8\over N_f} }. \cr } }
In a similar way to what we discussed for $Spin(7)$,
a number of consistency checks
of the duality $G_2$ --- $SU(N_f-3)$
can be performed. The R-symmetry is anomaly free,
the 't Hooft anomaly matching conditions are satisfied, etc.
Consider integrating out flavors {}from the electric theory with $N_f$
flavors. When $N_f\ge 7$,
the effect on the dual is to higgs the gauge group $SU(N_f-3)\to
SU(N_f-4)$, with the dual quark $q^{N_f}$ becoming massive.
The case $N_f=6$ deserves more attention.
The dual $SU(2)$ theory that is obtained has six
doublets $q^0$ and $q^i$, $i=1,\ldots,5$. Its physics is that
of confinement. The correct degrees of freedom in the infrared
are the mesons $B^i=q^0q^i$ and $A^{ij}=q^iq^j$ along with $M_{ij}$.
Since by duality this describes the $G_2$ theory with 5 flavors,
we learn that the independent gauge invariant chiral operators
of $G_2$ are a 2-index symmetric tensor $M$ of $SU(N_f)$, a 3-index totally
antisymmetric tensor $A$ and a 4-index totally antisymmetric tensor $B$.
Define an antisymmetric tensor $V$
by $V^{ij} = A^{ij}$, $i,j=1,\ldots,5$, $V^{i6}=B^i$, $i=1,\ldots,5$.
The $SU(2)$ dynamics gives a contribution to the
superpotential of $\pf V=\epsilon_{ijklm}
 B^i A^{jk} A^{lm}$.
Integrating out $s$, which is massive, produces two more terms
$M_{ij} M_{kl} A^{ik} A^{jl} + M_{ij} B^i B^j$, and we
also add $\det M$, allowed by the symmetries, as in the $Spin(7)$ case
with $N_f=7$ discussed above.
The resulting superpotential
\eqn\gtwofive{\det M + M_{ij}M_{kl}A^{ik}A^{jl} +
M_{ij} B^iB^j + \epsilon_{ijklm} B^i A^{jk} A^{lm} }
describes the confining $G_2$ theory with 5 flavors
whose equations of motion give the following
constraints on the expectation values
of $M, A$ and $B$:
\eqn\gtwoconstraints{ \eqalign{
\det M (M^{-1})^{ij}
+ M_{kl} A^{ik} A^{jl} + B^iB^j = 0 \qquad {\rm and}  \cr
M_{ik} M_{jl} A^{kl} +
\epsilon_{ijklm} B^{k} A^{lm} = 0 \qquad M_{ij} B^j +
\epsilon_{ijklm} A^{jk} A^{lm} = 0. \cr} }
Adding $m M_{55}$ to \gtwofive\ to integrate out the fifth flavor,
the equation of motion for $M_{55}$ gives a constraint
$ \det M^{(4)} + M^{(4)}_{ij} A^{(4)i} A^{(4)j} + B^{(4)}B^{(4)}
 = m$, where the label ${(4)}$ refers to the subset of
fields that belong to the effective theory
with 4 flavors. We learn {}from this constraint that
the quantum moduli space is smooth.
Keeping $M_{55}$ as a Lagrange multiplier,
we obtain the superpotential
$W_{(4)} = M_{55} ( \det M^{(4)} + M^{(4)}
 A^{(4)} A^{(4)} + B^{(4)} B^{(4)} + m)$.
Integrating out the fourth quark by adding $mM_{44}$,
we obtain a superpotential
$W_{(3)}= 1/ (\det M^{(3)} + (A^{(3)})^2)$.
Had we carefully kept track
of all the scales, we would have seen that this superpotential
has the correct power of the dynamical scale of the electric theory
to be generated by instantons.
Integrating out the two other flavors
lead to superpotentials
$W_{(i)}= 1/ (\det M^{(i)})^{1/(4-i)}$, $i=1,2$.
It is generated by gluino condensation in
the $SU(3)$ subgroup of $G_2$ for $i=1$ and in $SU(2)\in SU(3)$
for $i=2$. For $i=1,2,3$, there is no vacuum state.
Integrating out all the quarks
result in gluino condensation in the pure $G_2$ gauge theory.

As for future directions,
it would be especially interesting, by going up a few more steps, to reach
the $SO(10)$ gauge theories
with spinors, where the issues of grand unification, duality
and dynamical supersymmetry breaking possibly meet.

After the completion of this work, we learned of references
\refs{\ed - \giddings} which also examined some aspects of $G_2$ gauge
theories.

\centerline{{\bf Acknowledgments}}

We would like to thank V. Brazhnikov, K. Intriligator, N. Seiberg
 and M. Strassler for useful discussions,
and especially N. Seiberg for very valuable comments on the manuscript.
 This work was supported in part by DOE grant \#DE-FG05-90ER40559
and by a Canadian 1967 Science fellowship.
\listrefs
\bye